\documentclass[1p]{elsarticle}
\usepackage{bm}
\usepackage{amsmath,amssymb}
\usepackage{graphicx}

\newcommand{\nn}{\nonumber}

\newcommand{\be}{\begin{equation}}
\newcommand{\ee}{\end{equation}}
\newcommand{\ba}{\begin{align}}
\newcommand{\ea}{\end{align}}
\newcommand{\BM}{\begin{bmatrix}}
\newcommand{\EM}{\end{bmatrix}}

\newcommand{\bra}
   [1]{\mbox{$\langle #1|$}}
\newcommand{\ket}
   [1]{\mbox{$| #1\rangle$}}
\newcommand{\braket}
   [2]{\mbox{$\langle #1| #2\rangle $}}
\newcommand{\dbra}
   [1]{\mbox{$\bigl\langle\!\!\bigl\langle #1 \bigr|$}}
\newcommand{\dket}
   [1]{\mbox{$\bigl| #1\bigr\rangle\!\!\bigr\rangle$}}
\newcommand{\dbraket}
   [2]{\mbox{$\bigl\langle\!\!\bigl\langle #1\big| #2\bigr\rangle\!\!\bigr\rangle $}}
\newcommand{\Real}{\mathrm{Re}}
\newcommand{\Imag}{\mathrm{Im}}

\begin{document}
\begin{frontmatter}

\title{From superoperator formalism to nonequilibrium Thermo Field Dynamics}

\author[densi]{Y.~Nakamura}
\ead{nakamura@aoni.waseda.jp}
\author[densi]{Y.~Yamanaka}
\ead{yamanaka@waseda.jp}
\address[densi]{Department of Electronic and Photonic Systems, Waseda
University, Tokyo 169-8555, Japan}

\begin{abstract}
Emphasizing that the specification of the representation space or the 
quasiparticle picture is essential in nonequilibrium quantum field system, 
we have constructed the unique unperturbed representation of the interaction picture
in the superoperator formalism.  To achieve it, we put the three basic 
requirements (the existence of the
quasiparticle picture at each instant of time, the macroscopic causality and
the relaxation to equilibrium). From the resultant representation
follows the formulation of nonequilibrium Thermo Field Dynamics (TFD). 
The two parameters, the number distribution and excitation energy, characterizing
the representation, are to be determined by the renormalization condition. While
we point out that the diagonalization condition by Chu and Umezawa is inconsistent
with the equilibrium theory, we propose a new renormalization condition as a generalization
of the on-shell renormalization on the self-energy which derives the quantum transport equation
and determines the renormalized excitation energy.
\end{abstract}

\begin{keyword}
superoperator formalism \sep Thermo Field Dynamics 
\sep thermal field theory \sep nonequilibrium 
\sep transport equation \sep kinetic equation
\PACS 03.75.Kk \sep 05.70.Ln \sep 05.30.Jp
\end{keyword}

\end{frontmatter}

\section{Introduction}
A choice of the representation space, or the Fock space, is essential in quantum field theory. 
Without specifying it , it would be impossible to calculate matrix elements of 
field operators, namely physical quantities.
In quantum field theory of vacuum, we can rely on the picture of the stable 
asymptotic particle on the stable and unique vacuum, and the representation space
is the Fock space associated with the asymptotic particles.
Because the concept of the asymptotic field is achieved by the disappearance of 
 overlap of the wave-packets describing the particles in the infinite past and future, 
it is clearly invalid in the thermal system. 
If we insist on the existence of the asymptotic particles in thermal system, 
the S-matrix is unity (so-called ``trivial S-matrix") \cite{NRT,Landsman}.   
To this one may add the physical argument that a Hamiltonian for thermal system
is not bounded below (see e.g. Eq.~(\ref{eq:hatHS})) and that a stable particle picture
cannot be admitted, which is seen from non-vanishing imaginary part of the 
self-energy under the thermal situation. Hence every particle becomes thermally unstable
and should be called a quasiparticle. What we aim in this paper is to 
construct the quasiparticle picture explicitly in the unperturbed representation
of the interaction picture under thermal situation, which supplants the asymptotic particle
 picture in the vacuum theory.

There are known two nonequilibrium thermal field theories, i.e., 
the closed time path (CTP) formalism \cite{CTP} and Thermal Field Dynamics (TFD) \cite{AIP}. 
While the CTP formalism is widely used, 
we have been employing the TFD formalism because the concept of the representation space is clear even in nonequilibrium situation. 
In TFD, which is a real-time canonical formalism of quantum field theory, 
 the thermal degree of freedom is introduced through doubling each degree of freedom, 
and the mixed state expectation in the density matrix formalism is 
replaced with an average of a pure state vacuum, called the thermal vacuum. 
A well-defined quasiparticle picture is constructed on the doubled Fock space, 
and quasiparticle operators which diagonalize the unperturbed TFD Hamiltonian is defined in a self-consistent manner.
A time-dependent number distribution is introduced as an unknown parameter, 
and the self-consistent renormalization condition \cite{Chu2} derives an equation for it, 
i.e.~the quantum transport equation, which reduces to the well-known quantum Boltzmann equation in the Markovian limit.

So far we have investigated the cold atomic gas systems without and with Bose--Einstein condensate \cite{Cornell,Ketterle,Bradley}, 
and have derived the quantum transport equation in nonequilibrium TFD \cite{NSMOY,NakamuraYamanaka1}. 
The appropriate choice of the quasiparticle picture was essential there. 
As an important result of our previous works, our transport equation in the presence of 
Bose-Einstein condensate obtains an additional collision term, called the triple production term, 
which is absent in the other approaches \cite{Stoof,Zaremba,Gardiner2}.
While the triple production term vanishes for the stable condensate, it remains non-vanishing 
in case of the thermally unstable condensate
to prevent the system from equilibrating. 
Such instability, called the Landau instability and caused by collisions in 
which the negative-energy quasiparticles participate, 
has been observed in the cold atomic system \cite{Raman}. 
An approach using an inappropriate particle picture, 
for example the particle picture of original atom with energy spectrum $E(\bm{x},\bm{p}) = {p^2}/{2m} + V(\bm{x})$, 
takes account of no negative-energy particle and is inadequate to describe the Landau instability.
This is an example showing the importance of the choice of the quasiparticle picture.

In this paper we take a position that a representation space corresponding 
to a particular quasiparticle picture is chosen by taking an unperturbed representation
in the interaction picture. This is done in the superoperator formalism \cite{Schmutz}, following 
from the density matrix formalism. Then we put the three basic requirements which thermal 
field theory with the coexistence of microscopic and time-dependent 
macroscopic quantities is desired to
fulfill, and acquire the unique unperturbed representation
in which the unperturbed time-dependent density matrix has the geometrical
distribution as the equilibrium one. The discussions of this paper
 is confined to cases of no spontaneous breakdown of symmetry, i.e. without condensates.
 As will be easily seen,
the formulation of nonequilibrium TFD is derived from this representation.
It should be emphasized that what we have done in this paper is not a mere re-derivation of 
nonequilibrium TFD, but provides a new perspective and understanding of it. 
First, it is shown that the formulation of nonequilibrium TFD using the thermal vacuum and
time-dependent thermal Bogoliubov transformation, which has been assumed without a sound rationale, 
corresponds to the unique consistent choice of the unperturbed representation. We remark that 
the Feynman diagram method, a very powerful tool of quantum field theory though it is never
guaranteed in thermal field theory for nonequilibrium system, is available owing to it. 
Second, we point out an undesirable property of the renormalization condition 
proposed by Chu and Umezawa \cite{Chu2}, called the diagonalization condition. 
Explicitly the diagonalization condition is consistent only in the leading order, 
since it conflicts with the equilibrium theory in higher orders.
In the context of the discussions in the present paper, 
we find and propose a new self-consistent renormalization condition which overcomes the difficulty.
Our renormalization condition reduces to the diagonalization one in the leading order, 
nevertheless it is consistent with the equilibrium theory even in higher orders.

This paper is organized as follows. 
We briefly review the formulation of the superoperator formalism in Section 2. 
In Section 3, the Liouville--von Neumann equation is introduced in terms of superoperator formalism, 
and the Schr\"odinger, Heisenberg and interaction pictures are introduced in the Liouville space. 
In Section 4, the unperturbed quasiparticle representation is constructed from the three 
basic requirements.
The unique structure of the unperturbed density matrix is 
derived from the requirements in Section 5, 
which has been just an assumption in nonequilibrium TFD so far.
The equivalence of the superoperator formalism to nonequilibrium TFD is shown in Section 6. 
In Section 7, a new self-consistent renormalization condition is proposed.
Section 8 is devoted to summary and discussions.

\section{Superoperator formalism}

We briefly review the superoperator formalism for our discussions
 below, following Schmutz \cite{Schmutz}.

Let a quantum system be described by a set of annihilation- and 
creation-operators,
\be
[ a_j\,,\,  a_k^\dagger]_\sigma=\delta_{jk} \,,\qquad 
[ a_j\,,\,  a_k]_\sigma=[ a_j^\dagger\,,\,  a_k^\dagger]_\sigma=0
\ee
with
\begin{align}
[ A\,,\,  B]_\sigma &= AB-\sigma B A \nn \,,\\
\sigma & = 
\begin{cases}
1& \text{for bosons}\\
-1& \text{for fermions} \,.\\
\end{cases}
\end{align}
The Hilbert space (Fock space), denoted by ${\mathcal H}$,
 is spanned by the complete orthonormal set of basic vectors which are
 generated from
operations of $a_j^\dagger$ on the vacuum $\ket{0}$\,,
\begin{align}
\ket{m} &= \ket{m_1,m_2,\cdots}= \prod_j 
\frac{1}{\sqrt{m_j!}}\left(a_j^\dagger\right)^{m_j} \ket{0}\, ,\\
& \braket{m}{m'} =\delta_{mm'}\, , \qquad \sum_m \ket{m}\bra{m}=I \, .
\label{eq:ONCH}
\end{align}

The linear operators acting on ${\mathcal H}$ form
a linear space, called the Liouville space
 ${\bar {\mathcal H}}$\,.
We denote its element
by a double-ket $\dket{A}$, corresponding to the operator $A$ 
acting on ${\mathcal H}$,
where the notation of the double-ket $\dket{\cdot}$ is used to 
distinguish it from the vector $\ket{\cdot}$ of ${\mathcal H}$.
The inner
product of two elements is defined by
\be	\label{eq:defIP}
\dbraket{A}{B} = {\rm Tr}[{A^\dagger B}] \, .
\ee 
The complete orthonormal basis of ${\bar {\mathcal H}}$ is 
$\{\dket{m,n}= \dket{\ket{m}\bra{n}} \}$\,,
\be
\dbraket{m,n}{m',n'} =\delta_{mm'}\delta_{nn'}\, ,
 \qquad \sum_{m,n} \dket{m,n}\dbra{m,n}={\hat I} \,,
\label{eq:ONCbarH}
\ee 
where $\dbra{m,n} =\dbra{\ket{m}\bra{n}}$\,. For each operator acting $A$ on ${\mathcal H}$,
one has a superstate $\dket{A}$,
\be
\dket{A}= \sum_{m,n} A_{mn} \dket{m,n} \,,
\ee
with $A_{mn} = \dbraket{m,n}{A}= \bra{m}A\ket{n}$\,.

We introduce a concept of superoperator, operating on the
superstates, and a special set of superoperators, ${\check a}_j$ and 
${\tilde a}_j$, by
\be
{\check a}_j \dket{m,n}= \dket{a_j\ket{m}\bra{n}}\,,\qquad
{\tilde a}_j \dket{m,n}= \sqrt{\sigma} \sigma^{\mu -\nu} 
\dket{\ket{m}\bra{n}a_j^\dagger} \, ,
\label{eq:atilamn}
\ee
where $\mu=\sum_j m_j$ and $\nu=\sum_j n_j$\,. This phase convention of
the fermionic tilde operator is different from Schmutz's one \cite{Schmutz},
but is adopted from the references \cite{Henning,Chu2,Dzhioev,Chu3}. 
From the Hermitian conjugates of Eq.~(\ref{eq:atilamn}), 
\be
	\dbra{m,n} \check{a}_j^\dagger = \dbra{a_j\ket{m}\bra{n}} \,,\qquad
	\dbra{m,n} \tilde{a}_j^\dagger = \sqrt{\sigma}\sigma^{\mu-\nu+1}\dbra{\ket{m}\bra{n} a_j^\dagger} \,,
\ee
and the definition of the inner-product (\ref{eq:defIP}), one consistently obtains
\be
{\check a}_j^\dagger \dket{m,n}= \dket{a_j^\dagger\ket{m}\bra{n}}\,,\qquad
{\tilde a}_j^\dagger \dket{m,n}= \sqrt{\sigma} \sigma^{\mu -\nu} 
\dket{\ket{m}\bra{n}a_j} \, .
\label{eq:adtiladmn}
\ee
The superoperators ${\check a}_j$
and ${\tilde a}_j$ simulate the operations of $a_j$ on $\ket{m}\bra{n}$
 from the left and right, respectively.

It follows from  Eqs.~(\ref{eq:atilamn}) and (\ref{eq:adtiladmn}) that 
the superoperators obey the (anti)-commutation relations,
\begin{align}
& [{\check a}_j\,,\, {\check a}_k^\dagger]_\sigma
 = [{\tilde a}_j\,,\, {\tilde a}_k^\dagger]_\sigma=\delta_{jk}
\, , \nn\\ 
& [{\check a}_j^{(\dagger)}\,,\, {\tilde a}_k^{(\dagger)}]_\sigma=0
\, , \nn \\
& [{\check a}_j\,,\, {\check a}_k]_\sigma
=[{\check a}_j^\dagger\,,\, {\check a}_k^\dagger]_\sigma
=[{\tilde a}_j\,,\, {\tilde a}_k]_\sigma
=[{\tilde a}_j^\dagger\,,\, {\tilde a}_k^\dagger]_\sigma=0 \,.
\end{align}
The basic vectors of ${\bar {\mathcal H}}$ are generated from operations
of the creation superoperators on the supervacuum $\dket{0,0}$,
\be
\dket{m,n}=\left(\sigma \sqrt{\sigma} \right)^{\nu^2}
\biggl(\prod_{j'}\frac{1}{\sqrt{m_{j'}!}} \biggr)
\biggl(\prod_{k'}\frac{1}{\sqrt{n_{k'}!}} \biggr)
{\check a}_{j_1}^\dagger \cdots {\check a}_{j_\mu}^\dagger
{\tilde a}_{k_1}^\dagger \cdots {\tilde a}_{k_\nu}^\dagger
\dket{0,0} 
\, ,
\ee
and are the eigenstates of supernumbers,
\be
{\check a}_{j}^\dagger{\check a}_{j}\dket{m,n}=m_j \dket{m,n}
\,,
\qquad
{\tilde a}_{j}^\dagger {\tilde a}_{j}\dket{m,n}=n_j\dket{m,n}
\, .
\ee

The identity operator $I$ in (\ref{eq:ONCH}) gives the 
superstate $\dket{I}$,
\be
\dket{I} = \sum_m \dket{m,m} 
\,.
\ee
This, together with Eqs.~(\ref{eq:atilamn}) and (\ref{eq:adtiladmn}), yields
the relations,
\be
\left({\tilde a}_j -\sqrt{\sigma} {\check a}_{j}^\dagger \right)\dket{I}
= \left({\tilde a}_j^\dagger -
\sqrt{\sigma} {\check a}_{j} \right)\dket{I}=0
\, .
\label{eq:RTSI}
\ee

Suppose an operator $A=A(a^\dagger, a)$ in ${\mathcal H}$, then we define 
two superoperators by
\be
{\check A}= A({\check a}^\dagger, {\check a}) \, ,
\qquad {\tilde A}=A^\ast ({\tilde a}^\dagger, {\tilde a}) \,,
\ee
and refer to them as non-tilde and tilde operators, respectively.  
The relation between them is summarized  as the tilde conjugation rules,
\begin{align}
& \left({\check A}_1 {\check A}_2 \right)\Tilde{\phantom{i}} 
={\tilde A}_1 {\tilde A}_2 \,, \nn \\
&\left(c_1{\check A}_1+ c_2{\check A}_2 \right)\Tilde{\phantom{i}} 
=c_1^\ast {\tilde A}_1 + c_2^\ast {\tilde A}_2 \,, \nn
\\
& \left( {\check A}^\dagger\right)\Tilde{\phantom{i}} = {\tilde A}^\dagger \,,
\nn \\
& \left( {\tilde A}\right)\Tilde{\phantom{i}}= {\check A}\, ,
\label{eq:tilcon}
\end{align}
where $c_i$ are c-numbers. Note that the last property
and Eq.~(\ref{eq:RTSI}) are due to the phase convention for fermionic
 operator in Eq.~(\ref{eq:atilamn}) 
and differ from the conventional ones \cite{AIP}.
It can be shown from Eq.~(\ref{eq:RTSI}) that
\be
{\tilde A}^\dagger \dket{I} = 
\begin{cases}
{\check A}\dket{I} & \mbox{for bosonic $A$}\\
i{\check A} \dket{I} & \mbox{for fermionic $A$}\,.
\end{cases}
\label{eq:tilAdA}
\ee
The tilde conjugation may be applied to superstates, 
and we find
\be
\dket{m,n}\Tilde{\phantom{i}}=
 \left(\sqrt{\sigma} \right)^{(\mu+\nu)^2}\dket{n,m}
\, .
\ee
Therefore $\dket{I}$ becomes tilde-invariant,
\be
\dket{I}\Tilde{\phantom{i}}=\dket{I} \, .
\label{eq:tilI}
\ee
When $A$ is bosonic, we can show
\be
\dket{A}\Tilde{\phantom{i}} = 
\left({\check A}\dket{I} \right)\Tilde{\phantom{i}}
={\tilde A}\dket{I} = {\check A}^\dagger \dket{I}
=\dket{A^\dagger} \, ,
\ee
from Eqs.~(\ref{eq:tilI}) and (\ref{eq:tilAdA}).  This implies
that
\be
A^\dagger =A \quad \mbox{is equivalent to}\quad
\dket{A}\Tilde{\phantom{i}}= \dket{A} \quad
 \mbox{for bosonic $A$\,.}
\label{eq:AdagAtilde}
\ee

\section{Liouville--von Neumann equation in the Schr\"odinger, Heisenberg and interaction pictures}

Consider a quantum system under thermal situation in the Schr\"odinger picture, 
described by the density matrix $\rho_S(t)$.  
Its dynamical equation is the Liouville-von Neumann equation,
\be \label{eq:LvNeqS}
i  \frac{d}{dt}{ \rho}_S(t) 
= \left[H_S\, ,\, {\rho}_S(t) \right] \, ,
\ee
with the original  total Hamiltonian $H_S$\, which is Hermitian.
We take the unit of $\hbar=1$ throughout this paper.
We convert (\ref{eq:LvNeqS}) to the equation in the superoperator formalism, described
in the previous section:
\be
i  \frac{d}{dt} \dket{\rho_S(t)} ={\hat H}_S \dket{\rho_S(t)} \, ,
\ee
where the total super-Hamiltonian is 
\be
{\hat H}_S= {\check H}_S - {\tilde H}_S \,.
\label{eq:hatHS}
\ee

The Heisenberg picture is constructed as usual.
We introduce the time-evolution operator by
\be
{\hat U}(t,t_0) = \exp \left(-i
{\hat H}_S (t-t_0)\right) \, ,
\ee
and $t_0$ is the coincidence-time of the various pictures.
Then the time-independent state $\dket {\rho_H}$ is related to 
$\dket {\rho_S(t)}$ as
\be
\dket {\rho_S(t)}= {\hat U}(t,t_0) \dket {\rho_H} \, ,
\ee
while the Heisenberg superoperator is given by
\be
{\hat A}_H(t) = {\hat U}^{-1}(t, t_0) {\hat A}_S  {\hat U}(t, t_0) \, ,
\ee
and satisfies the Heisenberg equation,
\be
i  \frac{d}{dt} {\hat A}_H(t)  = \left[{\hat A}_H(t)\, ,\,
 {\hat H}({\check a}_H(t),\cdots)\right] \,.
\ee

In order to move to the interaction picture, the unperturbed 
super-Hamiltonian ${\hat H}_{\mathrm u}$ needs to be specified. 
Here are the two remarks regarding to it. Firstly it generally 
differs from the free super-Hamiltonian 
${\hat H}_0={\check H}_0-{\tilde H}_0$\,, constructed from quantum
field theory of vacuum. Secondly it involves some time-dependent parameters,
say $\eta(t)$, which will turn out to be the mean number $n_j(t)$
and the renormalized excitation energy $\omega_j(t)$,  and therefore depends 
explicitly on time, in contrast to the usual unperturbed Hamiltonian.
This is characteristic of our approach to
 describe time-dependent nonequilibrium 
processes, which will be clarified in later sections.
Once ${\hat H}_{\mathrm u}$ is specified, we introduce the time-dependent
superoperator ${\hat A}(t)$ in the interaction picture by
\be
{\hat A}(t) = {\hat U}_0^{-1}(t, t_0) {\hat A}_S  {\hat U}_0(t, t_0)
\ee
where
\be
i  \frac{d}{dt} {\hat U}_0(t, t_0)={\hat H}_{\mathrm u}
({\check a}_S,\cdots: \eta(t)) 
 {\hat U}_0(t, t_0) \, ,
\ee
and $\eta(t)$ inside ${\hat H}_{\mathrm u}$ represents 
time-dependent parameters,
mentioned above. The time-evolution operator ${\hat U}_0(t, t_0)$ is 
\be
{\hat U}_0(t, t_0)= {\rm T} \left[\exp \left(-i \int^t_{t_0}
{\hat H}_{\mathrm u}({\check a}_S,\cdots : \eta(s))\,ds \right) \right] \, ,
\label{eq:U0}
\ee
where the symbol ${\rm T}$ stands for a time-ordered product,
\be
{\rm T} \left[A(t_1)B(t_2) \right]=\theta(t_1-t_2) A(t_1)B(t_2)
+\sigma \theta(t_2-t_1)B(t_2) A(t_1) \, .
\ee
The equations for $\{{\check a}_j(t)\,,\,{\check a}_j^\dagger(t)
\,,\,{\tilde a}_j(t)\,,\,{\tilde a}_j^\dagger(t)\}$ are given by
\be
i  \frac{d}{dt} {\check a}_j(t) 
 = \left[{\check a}_j(t) , {\hat H}_{\mathrm u}(t)\right] \,,
\ee
and so on. Here the unperturbed super-Hamiltonian in the interaction picture
is denoted simply by ${\hat H}_{\mathrm u}(t)$\,,
\be
{\hat H}_{\mathrm u}(t)\equiv
{\hat H}_{\mathrm u}({\check a}(t),\cdots:\eta(t))= 
{\hat U}^{-1}_0(t, t_0) {\hat H}_{\mathrm u}
({\check a}_S,\cdots: \eta(t)) {\hat U}_0(t, t_0)\, .
\label{eq:HuiintHuS}
\ee
Note that the unperturbed Hamiltonians in both the interaction and
Schr\"odinger pictures do not coincide with each other in general 
due to the 
time-dependent parameters $\eta(t)$,
\be
{\hat H}_{\mathrm u}(t) \neq 
{\hat H}_{\mathrm u}({\check a}_S,\cdots: \eta(t)) \, .
\ee
The equal-time (anti)-commutation relations are
\be
\left[{\check a}_j(t)\,,  {\check a}_k^\dagger(t)\right]_\sigma
= \left[{\tilde a}_j(t)\,, 
 {\tilde a}_k^\dagger(t)\right]_\sigma=\delta_{jk}\, , 
\qquad \mbox{others}=0\, . 
\ee

The state in the interaction picture is
\begin{align}
\dket{\rho(t)}&= {\hat U}^{-1}_0(t, t_0) \dket{\rho_S(t)}
= {\hat V}(t,t_0)  \dket{\rho_S(t_0)} \, ,\nn
\\
& {\hat V}(t,t_0)= {\hat U}^{-1}_0(t, t_0) {\hat U}(t,t_0)
\,,  \nn \\
&i  \frac{d}{dt} {\hat V}(t, t_0)={\hat H}_{\mathrm I}(t) 
 {\hat V}(t, t_0) \, ,
\end{align}
with the interaction super-Hamiltonian,
\be
{\hat H}_{\mathrm I}(t)
= {\hat H}({\check a}(t),\cdots)-
{\hat H}_{\mathrm u}({\check a}(t),\cdots:\eta(t))
\, .
\label{eq:defhatHI}
\ee
Formally ${\hat V}(t, t_0)$ is
\be
{\hat V}(t, t_0)= {\rm T} \left[\exp \left(-i\int^t_{t_0}
{\hat H}_{\mathrm I}(s)\,ds \right) \right] \, .
\label{eq:defV}
\ee

The causal two-point Green's function is defined in the Heisenberg picture,
\be
G(t_1,t_2 )=-i \dbra{I}{\rm T}\left[
 {\hat A}_{H}(t_1){\hat B}_{H}(t_2)\right]
\dket{\rho_H}\, ,
\label{eq:2ptGFH}
\ee
which is $-i {\rm Tr} \left[ {\rm T}\left[A_H(t_1) B_H(t_2)\right]
 \rho_H \right]$ in the density matrix formalism, 
when ${\hat A}_{H}(t)={\check A}_H(t)$ and 
${\hat B}_{H}(t)={\check B}_{H}(t)$\,. We rewrite it in terms of 
the operators in the interaction picture,
putting $t_0=-\infty$,
\be
G(t_1,t_2 )=-i \dbra{I}{\hat S}^{-1}{\rm T}
\left[ {\hat S} {\hat A}(t_1){\hat B}(t_2)\right]
\dket{\rho_H} \, ,
\label{eq:2ptGFint}
\ee 
where the S-matrix superoperator is
\be
{\hat S} = {\hat V}(\infty, -\infty) \, .
\label{eq:Smatrix}
\ee

\section{Unperturbed representation}

In this section we attempt  to construct the unperturbed representation
in the superoperator formalism from general requirements for 
quantum field system under thermal situations.

We maintain that physical quantities for a quantum field system,
following from matrix elements of the field operators, 
can be calculated only when the Fock space representing
the abstract field operators is specified.  In quantum field theory
of vacuum, the existence of the stable and unique vacuum and the stable
 asymptotic particles makes the choice of the Fock space unambiguous, namely
 the Fock space of asymptotic fields is chosen. 
With $t_0=0$, the vacuum in the Heisenberg picture is related to that of
the asymptotic fields by the Gell-Mann--Low relation. Alternatively, 
one may take $t_0=-\infty$, then both of the vacua coincide with 
each other. In addition, because of $\bra{0}S^{-1}  \propto \bra{0}$ 
($S$: the S-matrix operator, $\ket{0}$: the asymptotic vacuum) 
we have the Feynman diagram method in calculating the causal Green's
functions.

The concepts of the vacuum as the ground state and the asymptotic
 particles are invalid in thermal situation. As is seen from 
Eq.~(\ref{eq:hatHS}), the total Hamiltonian ${\hat H}_S$ is not bounded below,
and there is no ground superstate.  The thermal situation does not allow
the asymptotic particles.  The requirement of the asymptotic particles
at finite temperature leads to the trivial S-matrix \cite{NRT,Landsman}.
Many works on thermal field theories respect the Heisenberg equation for 
the field operators, but do not make the representation space clear.
We will provide a definite representation space in the Liouville space,
constructing the unperturbed representation there, which follows from 
the very basic requirements.

Let us put the following three requirements.
\begin{description}
\item [(a)]  The unperturbed representation of the superoperator formalism
is constructed on the quasiparticle picture at each instant of time.
\item [(b)] In the causal Green's function, the macroscopic 
time-dependent quantities affect the microscopic motions only in the future
but not in the past, which we call the thermal causality. 
\item [(c)] The system approaches to equilibrium after a long time,
consistently with the law of thermodynamics.
\end{description}

The requirement (a) realizes the concept of the quasiparticle by Landau
in our treatment. We restrict ourselves in this paper to the case in which the original
Hamiltonian possess the global phase symmetry, and its spontaneous breakdown
or its rearrangement does not occur. (The present construction of the unperturbed
representation will be extended to cases with spontaneous breakdown of symmetry
or symmetry rearrangement in the future work. ) 
Consider the unperturbed system in the Schr\"odinger picture. Then (a) implies that
the unperturbed superstate $\dket{\rho_0(t)}$ exists for any $t$\,.  The superstate
$\dket{\rho_0(t)}$ which we seek is a special one, playing a role of ``vacuum" 
in the Liouville space on which the quasiparticle picture is constructed. 
As the global phase symmetry of the system is kept, we require
that $\dket{\rho_0(t)}$ should be
 invariant under the global phase transformation,
\be
{\check a}_j\, \rightarrow \, e^{i\theta} {\check a}_j\,, \qquad  
 {\tilde a}_j\, \rightarrow \, e^{-i\theta} {\tilde a}_j\,,
\label{eq:atilatransform}
\ee
and their Hermitian conjugates.  Then its general form is
\be
\dket{\rho_0(t)}=\sum_m p_m(t) \dket{m,m} 
\,.
\ee
The three basic properties of the density matrix are (i) the Hermiticity
$\rho_0^\dagger (t)=\rho_0 (t)$, 
(ii) the normalization ${\rm Tr}[\rho_0(t)]=1$ and (iii) the positivity
$ \bra{\phi} \rho_0(t) \ket{\phi} \geq 0$ for any $\ket{\phi}$\,, and
their expressions in terms of $p_m(t)$ are
\begin{alignat}{2}
p^\ast_m(t)& = p_m(t) \qquad& 	&\text{for (i)}\,,\\
\sum_m p_m(t)&= 1 & 			&\text{for (ii)}\,, \\
p_m(t) &\geq 0 	&				&\text{for (iii)}\,.
\end{alignat}
The mean unperturbed number $n_j(t)$ is given by
\be
n_j(t)= {\rm Tr}[a^\dagger_j a_j\rho_0(t)]= \dbra{I} {\check a}_j^\dagger
{\check a}_j \dket{\rho_0(t)}= \sum_m m_j p_m(t) \,.
\label{eq:defni}
\ee

The unperturbed super-Hamiltonian, which is diagonal for each mode, at most bilinear
in ${\check a}_j\,$, ${\check a}^\dagger_j\,$, ${\tilde a}_j$ 
and ${\tilde a}^\dagger_j$, and invariant under the transformation
in Eq.~(\ref{eq:atilatransform}), is generally expressed by
\begin{multline}
{\hat H}_{\mathrm u}({\check a},\cdots:\eta(t)) \\
= \sum_j\left\{ \eta_{1j}(t)\sigma \sqrt{\sigma} 
{\check a}_j {\tilde a}_j
+\eta_{2j}(t)  \sigma\sqrt{\sigma}{\check a}^\dagger_j {\tilde a}^\dagger_j
+\eta_{3j}(t) {\check a}_j^\dagger{\check a}_j
+\eta_{4j}(t)  {\tilde a}^\dagger_j {\tilde a}_j+ \eta_{5j}(t) \right\}\, ,
\label{eq:Hu1}
\end{multline}
where the operators are time-independent (the suffix $S$ is suppressed)
but ${\hat H}_{\mathrm u}({\check a},\cdots:\eta(t))$
becomes time-dependent through the five undetermined
parameters $\eta_{ij}(t)$ $(i=1,2, \cdots, 5)$ for each mode, complex in general. 
Note that ${\hat H}_{\mathrm u}({\check a},\cdots:\eta(t))$ is non-Hermitian. 
The requirement (a) also requires the unperturbed Liouville-von Neumann
equation,
\be
i  \frac{d}{dt} \dket{\rho_0(t)} =
{\hat H}_{\mathrm u}({\check a},\cdots:\eta(t)) \dket{\rho_0(t)} \, .
\label{eq:unLvN}
\ee

For the property (i) to be consistent with Eq.~(\ref{eq:unLvN}),
one has 
\be
\left\{{\hat H}_{\mathrm u}({\check a},\cdots:\eta(t))
\right\}\tilde{\phantom{i}}
=-{\hat H}_{\mathrm u}({\check a},\cdots:\eta(t))
\, ,
\label{eq:Hutilde}
\ee
as Eq.~(\ref{eq:AdagAtilde}) and the tilde conjugation of Eq.~(\ref{eq:unLvN}) show.
The condition (\ref{eq:Hutilde}) implies that the five complex $\eta_{ij}(t)$'s
are expressed by five real parameters for each mode \newline
$\{\omega_j(t),
\zeta_{1j}(t),\zeta_{2j}(t),\zeta_{3j}(t),\zeta_{5j}(t)\}$,
\begin{align}
\eta_{1j}(t)&= i \zeta_{1j}(t) \,, & \eta_{2j}(t) &= i \zeta_{2j}(t) \,, \nn \\
\eta_{3j}(t)&=   \omega_j(t) +i \zeta_{3j}(t)  \,,& 
\eta_{4j}(t)&= - \omega_j(t) +i\zeta _{3j}(t) \,, \nn \\ 
\eta_{5j}(t) &= i\zeta_{5j}(t)\, . &&  
\label{eq:cond1}
\end{align}

The conservation of probability (the property (ii)) is guaranteed
if 
\be
\dbra{I}{\hat H}_{\mathrm u}
({\check a},\cdots:\eta(t))=0 \,,
\label{eq:IHu}
\ee
which is seen from (\ref{eq:unLvN}). From this 
follow the two constraints
\be
\zeta_{1j}(t)+\sigma \zeta_{2j}(t)+2 \zeta_{3j}(t)=0\,, \qquad
\zeta_{2j}(t) +\zeta_{5j}(t)=0 \, .
\label{eq:cond2}
\ee

Using Eqs.~(\ref{eq:defni}) and (\ref{eq:unLvN}), one can calculate ${\dot n}_j(t)$ as
\be
{\dot n}_j(t)=\frac{1}{i } \dbra{I}{\check a}_j^\dagger
{\check a}_j{\hat H}_{\mathrm u}
({\check a},\cdots:\eta(t))\dket{\rho_0(t)}
= - \zeta_{1j}(t) n_j(t)+ \zeta_{2j}(t)
 \left(1+\sigma n_j(t) \right) \,. 
\label{eq:dotnzeta}
\ee

Equations (\ref{eq:cond2}) and (\ref{eq:dotnzeta}) form a 
simultaneous equations
for the four unknowns $\zeta_i(t)$ $(i=1,2,3,5)$ and their
 general solutions are
\begin{align}
&\zeta_{1j}(t) =\sigma {\dot n}_j(t) + \gamma_j(t)\, ,\quad
&\zeta_{2j}(t)&=-\zeta_{5j}(t)= {\dot n}_j(t) + \frac{n_j(t)}{1+\sigma n_j(t)}(t) \gamma_j(t) \,, \nn\\
&\zeta_{3j}(t) =-\sigma {\dot n}_j(t) - \frac{1+2\sigma n_j(t)}{2(1+\sigma n_j(t))} \gamma_j(t)\, ,&&
\label{eq:zeta-1}
\end{align}
with an arbitrary function $\gamma_j(t)$\,. The parameter $\omega_j(t)$ is
also arbitrary at this stage.

Let us move to the Heisenberg picture for the unperturbed system, or to
the interaction picture with the time-independent superstate $\dket{\rho_0}=
\dket{\rho_0(t_0)}\,$.  The Heisenberg equation, 
equivalent to Eq.~(\ref{eq:unLvN}), is
\be
i \frac{d}{dt}{\hat A}(t)= \left[ {\hat A}(t)\, , 
{\hat H}_{\mathrm u}(t) \right] \, ,
\qquad {\hat A}(t)= {\hat U}_0^{-1}(t,t_0){\hat A}_S{\hat U}_0(t,t_0) \, ,
\label{eq:unHeisenberg}
\ee
where the definition (\ref{eq:U0}) has been used and
\begin{align}
{\hat H}_{\mathrm u}(t) &= \sum_j\biggl[
\omega_j(t) \left\{{\check a}_j^\dagger(t) {\check a}_j(t)-
{\tilde a}_j^\dagger(t) {\tilde a}_j(t) \right\}  \nn \\
&\qquad   + i
\left\{ \zeta_{1j}(t)\sigma \sqrt{\sigma} 
{\check a}_j(t) {\tilde a}_j(t) + 
\zeta_{2j}(t)  \sigma\sqrt{\sigma}{\check a}^\dagger_j(t) 
{\tilde a}^\dagger_j(t)\right. \nn \\
&   \left. \qquad \qquad \quad+ 
\zeta_{3j}(t)\left({\check a}_j^\dagger(t){\check a}_j(t)
+{\tilde a}^\dagger_j(t) {\tilde a}_j(t)\right)
+ \zeta_{5j}(t) \right\} \biggr] \,,
\label{eq:Hu2}
\end{align}
with Eq.~(\ref{eq:zeta-1}). 
%

Although the availability of the Feynman diagram method 
in  calculating the causal Green's functions seems a technical matter,
it is crucial.  Without the Feynman diagram method,
a systematic analysis of the Green's function would be
impossible, since we could not use the powerful tools 
of quantum field theory such as the Wick theorem, 
the Dyson--Schwinger equation, the concepts of one-particle
irreducible diagram and the effective action, 
the renormalization procedure and so on. For the availability
of the Feynman diagram, one should have 
$\dbra{I}{\hat S}^{-1} \propto \dbra{I}$\, 
in Eq.~(\ref{eq:2ptGFint}).  This relation holds if the 
interaction super-Hamiltonian
in Eq.~(\ref{eq:defhatHI}) satisfies
\be
\dbra{I}{\hat H}_{\mathrm I}(t)=0 \, .
\label{eq:bra1hatHI}
\ee
As it can be shown that $\dbra{I}{\hat H}(t)=0$ for the Hermitian $H$, 
we only have to check $\dbra{I}{\hat H}_{\mathrm u}(t)=0$\,. 
Equation (\ref{eq:IHu}), coming from the conservation of probability, is true
for arbitrary $n_j(t)$\,, $\gamma_j(t)$\, and $\omega_j(t)$, and 
${\hat H}_{\mathrm u}(t)$ is related to 
${\hat H}_{\mathrm u}({\check a}_S,\cdots: \eta(t))$ as 
in Eq.~(\ref{eq:HuiintHuS}) with Eq.~(\ref{eq:U0}), so we have
\be
\dbra{I}{\hat H}_{\mathrm u}(t)=0 \,.
\ee
Thus the Feynman diagram method is available.

The mean number $n_j(t)$ is a macroscopic quantity, given by
the expectation over the density matrix $\rho_0(t)$\,. It is expected that 
our unperturbed Hamiltonian  governs the microscopic quantum
dynamics in a consistent manner with the macroscopic 
thermal change in the background. The dependence of 
${\hat H}_{\mathrm u}(t)$ on $n_j(t)$ means that the operators 
in the interaction picture $\{{\check a}_j(t)\,,\,{\check a}_j^\dagger(t)
\,,\,{\tilde a}_j(t)\,,\,{\tilde a}_j^\dagger(t)\}$ depend on
$n_j(t)$\,. Suppose the two-point unperturbed causal Green's function
or the unperturbed propagator, 
\begin{multline}
\Delta (t_1,t_2) =-i \dbra{I} {\rm T} \left[ {\hat a}_1(t_1){\hat a}_2(t_2)
\right] \dket{\rho_0}\\
 = 
-i\theta(t_1-t_2) \dbra{I}  {\hat a}_1(t_1){\hat a}_2(t_2)\dket{\rho_0}
-i\sigma \theta(t_2-t_1)\dbra{I}  {\hat a}_2(t_2){\hat a}_1(t_1) 
\dket{\rho_0}\, ,
\label{eq:prop}
\end{multline}
where ${\hat a}_i$ stands for any of 
$\{{\check a}_j(t)\,,\,{\check a}_j^\dagger(t)
\,,\,{\tilde a}_j(t)\,,\,{\tilde a}_j^\dagger(t)\}$\,.
Generally speaking, $\Delta(t_1,t_2)$ depends on $t_1$ and $t_2$, and
on both of $n_j(t_1)$ {\em and} $n_j(t_2)$\,. The requirement (b) demands that
it should depend only on $n_j(t_2)$ but not on $n_j(t_1)$ for $t_1> t_2$
and vice verse for $t_2> t_1$\,. Otherwise the microscopic dynamics would
be affected by the macroscopic quantities in the future.

Let us see what constraint emerges when (b) is imposed. 
We have the three parameters $\omega_j(t)\,,\, n_j(t)\,,\, \gamma_j(t)$\,.
The parameter $\omega_j(t)$ is interpreted as the renormalized 
excitation energy and belongs to the microscopic  dynamics. On the other hand,
$n_j(t)$ and $\gamma_j(t)$, appearing in the imaginary part of
the super-Hamiltonian, represent thermal situation, and therefore are 
macroscopic parameters though the physical meaning of $\gamma_j(t)$ is
 not clear at this point. To extract the macroscopic part in the unperturbed Hamiltonian,
we introduce new operators 
$\{{\check \alpha}_j(t)\,,\,{\check \alpha}_j^\dagger(t)
\,,\,{\tilde \alpha}_j(t)\,,\,{\tilde \alpha}_j^\dagger(t)\}$\, 
from $\{{\check a}_j(t)\,,\,{\check a}_j^\dagger(t)
\,,\,{\tilde a}_j(t)\,,\,{\tilde a}_j^\dagger(t)\}$\,,
\begin{align}
&{\check a}_j(t) = e^{-i \int^t\!\omega_j(s)\,ds} {\check \alpha}_j(t)\,,
\qquad 
&{\tilde a}_j^\dagger(t) = e^{-i \int^t\!\omega_j(s)\, ds}
 {\tilde \alpha}_j^\dagger(t)\,,  \nn\\
&{\check a}_j^\dagger(t) = e^{i \int^t\!\omega_j(s)\,ds}
{\check \alpha}_j^\dagger(t)  \,,
&{\tilde a}_j(t)= e^{i \int^t\!\omega_j(s)\,ds}
 {\tilde \alpha}_j(t) \, .
\end{align}
The unperturbed super-Hamiltonian for the new operators is
\begin{align}
{\hat H}_{\alpha{\mathrm u}}(t) &= i \sum_j\left[
 \zeta_{1j}(t)\sigma \sqrt{\sigma} 
{\check \alpha}_j(t) {\tilde \alpha}_j(t) + 
\zeta_{2j}(t)  \sigma\sqrt{\sigma}{\check \alpha}^\dagger_j(t) 
{\tilde \alpha}^\dagger_j(t)\right. \nn \\
&   \left. \qquad\quad+ 
\zeta_{3j}(t)\left({\check \alpha}^\dagger(t){\check \alpha}(t)
+{\tilde \alpha}^\dagger_j(t) {\tilde \alpha}_j(t)\right)
+ \zeta_{5j}(t)  \right] 
\label{eq:Hualpha1}
\end{align}
with Eq.~(\ref{eq:zeta-1}), and their equal-time (anti)-commutation relations are
\be
\left[{\check \alpha}_j(t)\,,  {\check \alpha}_k^\dagger(t)\right]_\sigma
= \left[{\tilde \alpha}_j(t)\,, 
 {\tilde \alpha}_k^\dagger(t)\right]_\sigma=\delta_{jk}\, , 
\qquad \mbox{others}=0\, . 
\ee

Replace ${\hat a}_i(t)$ with ${\hat \alpha}_i(t)$ in Eq.~(\ref{eq:prop}) and 
consider  for $t_1>t_2$\,. As the time-dependence of ${\hat \alpha}_i(t)$
comes solely from $n_j(t)$ and $\gamma_j(t)$\,, it is sufficient for (b) that
$\dbra{I} {\hat \alpha}_1(t_1)$ is independent of $t_1$\,. 
Similarly for $t_2>t_1$
$\dbra{I} {\hat \alpha}_2(t_2)$ is independent of $t_2$\,.
Thus the requirement (b) is fulfilled if we have all of the following
relations:
\be
0=\dbra{I}\dot{\check \alpha}_j(t)
=\dbra{I}\dot{\check \alpha}_j^\dagger(t)
=\dbra{I}\dot{\tilde \alpha}_j(t)
=\dbra{I}\dot{\tilde \alpha}_j^\dagger(t) \,.
\label{eq:dotalphaI}
\ee
The equations for $\alpha(t)$'s are
\begin{align}
\dot{\check \alpha}_j(t) &=
\zeta_{2j}(t)  \sigma\sqrt{\sigma}{\tilde \alpha}^\dagger_j(t)
+ \zeta_{3j}(t){\check \alpha}_j(t)  \,,\nn \\
\dot{\check \alpha}_j^\dagger(t)&=
-\zeta_{1j}(t)  \sqrt{\sigma}{\tilde \alpha}_j(t)
- \zeta_{3j}(t){\check \alpha}_j^\dagger(t)  \,,\nn \\
\dot{\tilde \alpha}_j(t)&= 
\zeta_{2j}(t)  \sqrt{\sigma}{\check \alpha}^\dagger_j(t)
+ \zeta_{3j}(t){\tilde \alpha}_j(t)  \,,\nn \\
\dot{\tilde \alpha}_j^\dagger(t)&=
-\zeta_{1j}(t) \sigma \sqrt{\sigma}{\check \alpha}_j(t)
- \zeta_{3j}(t){\tilde \alpha}_j^\dagger(t)  \,, 
\label{eq:eqalpha1}
\end{align}
from the Heisenberg equations $i \dot{\check \alpha}_j(t)
 =[{\check \alpha}_j(t)\,,\, {\hat H}_{\alpha{\mathrm u}}(t)]$\,
 and so on.
We also note
\be
\dbra{I}\left({\tilde \alpha}_j (t)-\sigma
\sqrt{\sigma} {\check \alpha}_{j}^\dagger(t) \right)
= \dbra{I}\left({\tilde \alpha}_j^\dagger(t) -
\sigma\sqrt{\sigma} {\check a}_{j}(t) \right)=0
\, ,
\label{eq:LTSIalpha}
\ee
derived from Eq.~(\ref{eq:RTSI}) and $\dbra{I}{\hat H}_{\alpha{\mathrm u}}(t)=0$\,.
We find from Eqs.~(\ref{eq:eqalpha1}) and (\ref{eq:LTSIalpha}) that 
Eq.~(\ref{eq:dotalphaI}) holds true only if $\sigma \zeta_{2j}(t)+\zeta_{3j}(t)=0$
and $\zeta_{1j}(t)+\zeta_{3j}(t)=0$ or equivalently 
\be
\gamma_j(t)=0\,.
\label{eq:gamma0}
\ee

This way the unperturbed representation can be described by only the
two physical parameters $\omega_j(t)$ and $n_j(t)$,
which reflect thermal situations of the system and can be adjusted 
freely to them. Explicitly the
unperturbed Hamiltonian in the interaction picture is
\begin{align}
{\hat H}_{\mathrm u}(t) &= \sum_j\left[
\omega_j(t) \left\{{\check a}_j^\dagger(t) {\check a}_j(t)-
{\tilde a}_j^\dagger(t) {\tilde a}_j(t) \right\} \right. \nn \\
&\qquad \left.  + i{\dot n}_j(t)
\left\{ \sqrt{\sigma} 
{\check a}_j(t) {\tilde a}_j(t) + 
 \sigma\sqrt{\sigma}{\check a}^\dagger_j(t) 
{\tilde a}^\dagger_j(t)
-\sigma {\check a}_j^\dagger(t){\check a}_j(t)
-{\tilde a}_j(t) {\tilde a}_j^\dagger(t) \right\} \right] \,,
\label{eq:Hu3}
\end{align}
or
\begin{align}
{\hat H}_{\alpha{\mathrm u}}(t) &= i\sum_j {\dot n}_j(t)
\left\{ \sqrt{\sigma} 
{\check \alpha}_j(t) {\tilde \alpha}_j(t) + 
 \sigma\sqrt{\sigma}{\check \alpha}^\dagger_j(t) 
{\tilde \alpha}^\dagger_j(t)
-\sigma {\check \alpha}_j^\dagger(t){\check \alpha}_j(t)
-{\tilde \alpha}_j(t) {\tilde \alpha}^\dagger_j(t) \right\} 
\nn \\
&= -i\sum_j {\dot n}_j(t)
\left\{ {\tilde \alpha}_j(t)-\sigma\sqrt{\sigma}{\check \alpha}^\dagger_j(t) \right\}
\left\{{\tilde \alpha}^\dagger_j(t)-\sigma\sqrt{\sigma}{\check \alpha}_j(t) 
\right\}\,.
\label{eq:Hualpha2}
\end{align}

\section{Derivation of geometric distribution} 

We study the unperturbed dynamics obtained in the previous section.

Substituting Eq.~(\ref{eq:gamma0}) into Eq.~(\ref{eq:zeta-1}) and rewriting
Eq.~(\ref{eq:eqalpha1}), we have
\begin{align}
\dot{\check \alpha}_j(t) &=
-\sigma {\dot n}_j(t) 
\left\{{\check \alpha}_j(t) -\sqrt{\sigma}{\tilde \alpha}^\dagger_j(t)
\right\}  \,,\nn \\
\dot{\tilde \alpha}_j^\dagger(t)&=
-\sqrt{\sigma} {\dot n}_j(t) 
\left\{{\check \alpha}_j(t) -\sqrt{\sigma}{\tilde \alpha}^\dagger_j(t)
\right\}  \,,\nn \\
\dot{\tilde \alpha}_j(t)&= 
-\sigma {\dot n}_j(t) 
\left\{{\tilde \alpha}_j(t) -\sigma \sqrt{\sigma}{\check \alpha}^\dagger_j(t)
\right\}  \,,\nn \\
\dot{\check \alpha}_j^\dagger(t) &=
-\sigma\sqrt{\sigma} {\dot n}_j(t) 
\left\{{\tilde \alpha}_j(t) -\sigma \sqrt{\sigma}{\check \alpha}^\dagger_j(t)
\right\}  \,. 
\label{eq:eqalpha2}
\end{align}
It is easy to see that
\be
\frac{d}{dt}
 \left\{{\check \alpha}_j(t) -\sqrt{\sigma}{\tilde \alpha}^\dagger_j(t) \right\}
=\frac{d}{dt}
\left\{{\tilde \alpha}_j(t) -\sigma \sqrt{\sigma}{\check \alpha}^\dagger_j(t)
\right\} =0 \,.
\label{eq:conserve1}
\ee
This can be understood from the fact that 
${\check \alpha}_j(t) -\sqrt{\sigma}{\tilde \alpha}^\dagger_j(t)$
and its tilde conjugate commute with ${\hat H}_{\alpha{\mathrm u}}(t)$ in 
Eq.~(\ref{eq:Hualpha2}).  

We also have the following time-independent combinations,
\be
\frac{d}{dt}
 \left\{(1+\sigma n_j(t)){\check \alpha}_j(t)
 -\sigma\sqrt{\sigma} n_j(t){\tilde \alpha}^\dagger_j(t) \right\}
=\frac{d}{dt}
\left\{(1+\sigma n_j(t)){\tilde \alpha}_j(t) -
\sqrt{\sigma}  n_j(t) {\check \alpha}^\dagger_j(t)
\right\} =0 \,.
\label{eq:conserve2}
\ee
The proof of the first one is as follows:
\begin{multline}
\frac{d}{dt}
 \left\{(1+\sigma n_j(t)){\check \alpha}_j(t)
 -\sigma\sqrt{\sigma}  n_j(t){\tilde \alpha}^\dagger_j(t) \right\}
=\sigma  n_j(t)\frac{d}{dt}
 \left\{{\check \alpha}_j(t) -\sqrt{\sigma}{\tilde \alpha}^\dagger_j(t) \right\} \\
+ \dot{\check \alpha}_j(t)+\sigma  {\dot n}_j(t) \left\{ {\check \alpha}_j(t) 
-\sqrt{\sigma}{\tilde \alpha}^\dagger_j(t)\right\}  
= 0 \,,
\end{multline}
due to Eqs.~(\ref{eq:eqalpha2}) and (\ref{eq:conserve1}). The second one
 can be proven similarly. 

We shall elicit the implications of Eq.~(\ref{eq:conserve2}) 
for boson ($\sigma=1$). For simplicity, we here 
discuss a model of a single mode,
but the result can readily be generalized to multi-mode cases.
Noting that
\begin{align}
{\check \alpha}(t) &= {\hat U}_{\alpha 0}^{-1}(t,t_0) {\check a}
{\hat U}_{\alpha 0}(t,t_0) \, , \\
{\hat U}_{\alpha 0}(t,t_0) &=
{\rm T} \left[\exp \left(-i \int^t_{t_0}
{\hat H}_{\alpha{\mathrm u}}({\check a},\cdots : n(s))\,ds \right) \right] \,,
\end{align}
and
\be
\dket{\rho_0(t)} = {\hat U}_{\alpha 0}(t,t_0)\dket{\rho_0}= 
\sum_m p_m(t) \dket{m,m} \,.
\ee

Operate $\displaystyle \left\{\bigl(1+n(t)\bigr){\check \alpha}(t) - 
 n(t){\tilde \alpha}^\dagger(t)\right\}$
on $\dket{\rho_{0}}$, then we have
\begin{multline}
\left\{\bigl(1+n(t)\bigr){\check \alpha}(t) -  n(t){\tilde \alpha}^\dagger(t)
\right\} \dket{\rho_{0}} \\
= {\hat U}_{\alpha}^{-1} (t,t_0)
\sum_{m} \sqrt{m+1} \left\{\bigl(1+n(t)\bigr)p_{m+1}(t)-n(t) p_{m}
(t) \right\} 
\dket{m+1,m} \, .
\end{multline} 
Equation (\ref{eq:conserve2}) implies for arbitrary $t_1$ and $t_2$,
\begin{align}
&& \sum_m q_m(t_1)
\dket{m+1,m}=\sum_m q_m(t_2){\hat U}_\alpha(t_1,t_2)
\dket{m+1,m}\, , 
\label{eq:conserve22}\\
&& q_m(t)= \sqrt{m+1}\left\{\bigl(1+n(t)\bigr)p_{m+1}(t)-n(t) p_m(t)\right\} \qquad (m=0,1,2,\cdots) 
\, .
\end{align}
Putting $t_1=t+\Delta t\, , \, t_2=t$ and taking the limit 
$\Delta t \rightarrow 0$, we derive
\begin{align}
{\dot q}_m(t)={\dot n}(t) \sqrt{m+1}
\left\{\sqrt{m+2}\; q_{m+1}(t)  -2\sqrt{m+1}\; q_m(t)   +\sqrt{m}\; q_{m-1}(t) \right\}
\nn \\
 (m=0,1,2,\cdots) \,,
\label{eq:qmeq}
\end{align}
or equivalently
\be
	\dot{\bm q}(t) = \dot{n}(t) M \bm q(t) \,,
\label{eq:qmeq_vector}
\ee
with a real symmetric matrix $M$
\be
	M_{mm'} = \begin{cases}
	-2(m+1)& (m=m')\\
	\sqrt{(m+1)\,(m'+1)} \quad& (m=m'\pm1) \\
	0 & (\text{otherwise})
	\end{cases}
	\qquad
	(m=0,1,2,\cdots) 
	\,.
\ee
Any $\bm q$ at arbitrary time $t$ can be expanded as
\be
	\bm q(t) = \sum_\ell c_\ell(t) \bm u_\ell \,,
\ee
with the orthonormal-complete set of eigenvectors, $\{\bm u_\ell\}$\,:
\be
	M \bm u_\ell = \lambda_\ell \bm u_\ell \,.
\ee

Solving Eq.~(\ref{eq:qmeq_vector}), we obtain
\be
	\bm{q}(t) = \sum_\ell c_\ell(\tau)\bm u_\ell\, 
e^{\lambda_\ell \bigl[n(\tau) - n(t)\bigr]} \,,
\ee
which implies
\be
	\| \bm{q} (t)\|^2 = \sum_\ell |c_\ell(\tau)|^2 \,
e^{2\lambda_\ell \bigl[n(\tau) - n(t)\bigr]} \,.
\label{eq:qnorm2}
\ee

The trivial solution of Eq.~(\ref{eq:qmeq_vector}) is 
\be
	\bm q(t)=\bm0\,,
	\label{eq:solq}
\ee
or equivalently  $c_\ell(t) = 0$ for all $\ell$, 
which in turn leads to the geometrical distribution for $p_m(t)$
\begin{align}
&\frac{p_{m+1}(t)}{p_m(t) }= \frac {n(t)}{1+n(t)}=f(t)
 \qquad \mbox{(independent of $m$)} \nn \,,\\
&p_m(t)= (1-f(t)) f^m(t)\, .
\end{align}
Equation (\ref{eq:qnorm2}) indicates 
that if $\bm q(\tau)\ne\bm0$ at a certain time $\tau$ then $\bm q(\infty)\ne\bm0$. 
 If we require the approach to 
equilibrium (the requirement (c)) for which $\bm q_{{\mathrm {eq}}} = \bm0$\,,
then only the meaningful solution is Eq.~(\ref{eq:solq}) and $p_m(t)$
is the geometrical distribution,
\be
\dket{\rho_0(t)} = (1-f(t)) \sum_m f^m(t) \dket{m,m} \,.
\label{eq:GeoDisBose}
\ee

For the fermionic case, there are only $p_0(t)$ and $p_1(t)$, so 
we always write
\be
\dket{\rho_0(t)} = (1+f(t))^{-1} \sum_m f^m(t) \dket{m,m} \,.
\label{eq:GeoDisFermi}
\ee
Equations (\ref{eq:GeoDisBose}) and (\ref{eq:GeoDisFermi}) are unified into
a single expression,
\be
\dket{\rho_0(t)} = (1-\sigma f(t))^{\sigma} \sum_m f^m(t) \dket{m,m} \,.
\label{eq:GeoDis}
\ee

Thus for the time-independent superstate $\dket{\rho_0}$, consistent with the requirement
(c), we have
\begin{align}
&\left\{(1+\sigma n_j(t)){\check \alpha}_j(t)
 -\sigma\sqrt{\sigma} n_j(t){\tilde \alpha}^\dagger_j(t) \right\}\dket{\rho_0}
\nn\\ =
&\left\{(1+\sigma n_j(t)){\tilde \alpha}_j(t) -
\sqrt{\sigma} n_j(t) {\check \alpha}^\dagger_j(t)
\right\}\dket{\rho_0} 
=0 \,.
\label{eq:TSCrho0}
\end{align}

\section{Equivalence to nonequilibrium Thermo Field Dynamics}

We rewrite the superoperator formalism, obtained from
the basic requirements, in language of Thermo Field
Dynamics \cite{AIP}. The operators $\{ {\check a}_j(t)\,,\, 
{\check a}_j^\dagger (t)\}$ and $\{ {\tilde a}_j(t)\,,\, 
{\tilde a}_j^\dagger (t)\}$ in the interaction picture
can be identified as the non-tilde and tilde ones
in the nonequilibrium TFD formalism, which form the thermal doublets,
\be
a^\mu_j(t)=
\begin{bmatrix}
{\check a}_j(t)\\
\sqrt{\sigma}{\tilde a}_j^\dagger (t) 
\end{bmatrix}
\, , \qquad 
{\bar a}^\mu_j (t)=
\begin{bmatrix}
{\check a}_j^\dagger(t)& 
-\sqrt{\sigma}{\tilde a}_j (t) 
\end{bmatrix}
\,, \qquad (\mu=1,2) \,,
\ee
with the equal-time (anti)-commutation relations,
\be
[ a^\mu_j(t)\, , \,{\bar a}^\nu_k (t)]_\sigma
 = \delta_{jk}\delta_{\mu\nu}
\, , \qquad [ a^\mu_j(t)\, , \,a^\nu_k(t)]_\sigma
 =[ {\bar a}^\mu_j (t)\, , \,{\bar a}^\nu_k (t)]_\sigma=0\,.
\ee
The superstates $\dbra{I}$ and $\dket{\rho_0}$
become the thermal vacua, denoted simply by $\bra{0}$ and $\ket{0}$,
for which the following thermal state conditions hold,
\begin{align}
&\bra{0}\left({\tilde a}_j (t)-\sigma
\sqrt{\sigma} {\check a}_{j}^\dagger(t) \right)
= \bra{0}\left({\tilde \alpha}_j^\dagger(t) -
\sigma\sqrt{\sigma} {\check a}_{j}(t) \right)=0
\, ,
\label{eq:BVTS}\\
&\left\{(1+\sigma n_j(t)){\check a}_j(t)
 -\sigma\sqrt{\sigma} n_j(t){\tilde a}^\dagger_j(t) \right\}
\ket{0}\nn \\
& =\left\{(1+\sigma n_j(t)){\tilde a}_j(t) -
\sqrt{\sigma}  n_j(t){\check a}^\dagger_j(t)
\right\}\ket{0} =0 \,.
\label{eq:KVTS}
\end{align}
from Eqs.~(\ref{eq:LTSIalpha}) and (\ref{eq:TSCrho0}).

The properties in Eqs.~(\ref{eq:BVTS}) and (\ref{eq:KVTS}) enable us to introduce
the time-independent \newline 
annihilation- and creation-operators associated with the
thermal vacua, 
$\{{\check \xi}_j\,,\,{\check \xi}_j^\dagger\,,\,{\tilde \xi}_j
\,,\,{\tilde \xi}_j^\dagger\}$\,,
\be
{\check \xi}_j\ket{0}= {\tilde \xi}_j\ket{0}=0\,,
\qquad
\bra{0}{\check \xi}_j^\dagger=\bra{0}{\tilde \xi}_j^\dagger
\,.
\ee
with
\begin{align}
\xi^\mu_j(t)&=
\begin{bmatrix}
{\check \xi}_j(t)\\
\sqrt{\sigma}{\tilde \xi}_j^\dagger  (t)
\end{bmatrix}
=  B^{\mu\nu}_j(t) a^\nu_j(t)\,,
\qquad
{\bar \xi}^\mu_j(t) =
\begin{bmatrix}
{\check \xi}_j^\dagger(t) & 
-\sqrt{\sigma}{\tilde \xi}_j (t)
\end{bmatrix}
={\bar a}^{\nu}_j(t)  B^{-1,\nu\mu}_j(t)
\,,
\label{eq:xiBa}
\end{align}
where
\begin{align}
&{\check \xi}_j(t) = e^{-i \int^t\!\omega_j(s)\,ds} {\check \xi}_j\,,\quad 
&{\tilde \xi}_j^\dagger(t) = e^{-i \int^t\!\omega_j(s)\,ds} {\tilde \xi}_j^\dagger\,,  \nn\\
&{\check \xi}_j^\dagger(t) = e^{i \int^t\!\omega_j(s)\,ds}{\check \xi}_j^\dagger  \,,
&{\tilde \xi}_j(t)= e^{i \int^t\!\omega_j(s)\,ds} {\tilde \xi}_j \,.
\end{align}
Note that the dummy thermal index is summed in what follows. 
The thermal Bogoliubov matrix $B_j^{\mu\nu}(t)$ is defined by
\be
B_j^{\mu\nu}(t)=
\begin{bmatrix}
1+\sigma n_j(t)& -\sigma n_j(t)\\
-1 & 1
\end{bmatrix}
\ee
and the (anti)-commutation relations of the $\xi$-operators are
\be
[ \xi^\mu_j\, , \,{\bar \xi}^\nu_k ]_\sigma
 = \delta_{jk}\delta_{\mu\nu}
\, , \qquad [ \xi^\mu_j\, , \,\xi^\nu_k]_\sigma
 =[ {\bar \xi}^\mu_j \, , \,{\bar \xi}^\nu_k ]_\sigma=0\,.
\ee
Equation (\ref{eq:xiBa}) can be inverted into
\be
a^\mu_j(t) =  B^{-1,\mu\nu}_j(t)\xi^\nu_j(t)
\,,\qquad
{\bar a}^{\mu}_j(t)= {\bar \xi}^\nu_j(t) B^{\nu\mu}_j(t)
\, ,
\ee
which are the starting relations in nonequilibrium TFD \cite{AIP}.

The unperturbed and interaction Hamiltonians are Eq.~(\ref{eq:Hu3})
and
\begin{align}
{\hat H}_{\mathrm I}(t) &= {\hat H}- {\hat H}_{\mathrm u}(t) \nn \\
& {\hat H}= {\hat H}_0+ {\hat H}_{\mathrm{int}} \nn \\
& {\hat H}_0= \sum_j \omega_{0j} \left\{{\check a}_j^\dagger(t) {\check a}_j(t)-
{\tilde a}_j^\dagger(t) {\tilde a}_j(t) \right\} \,,
\end{align}
respectively,
where ${\hat H}_{\mathrm{int}}= {\check H}_{\mathrm{int}}-  {\tilde H}_{\mathrm{int}}$
represents  a non-linear interaction.  Note that the counter terms, proportional 
to $\delta \omega_j(t)=\omega_{j}(t) -\omega_{0j}$ and ${\dot n}_j(t)$\,, are present
in ${\hat H}_{\mathrm I}$ and that they will be determined from the self-consistent
renormalization condition. In the doublet notation, ${\hat H}_{\mathrm u}(t)$ is
\be
{\hat H}_{\mathrm u}(t) = \sum_j\left[
\omega_j(t) \left\{{\bar a}_j^\mu(t) a^\mu_j(t)+\sigma\right\}
- i\sigma {\dot n}_j(t) {\bar a}_j^\mu(t)T_0^{\mu\nu} a^\nu_j(t)\right]
\,,
\label{eq:Hu4}
\ee
where 
\be
T_0=
\begin{bmatrix}
1& -1\\
1&-1
\end{bmatrix}
\,.
\ee

The causal Green's function for the Heisenberg operators is defined by
\be
G_{jk}^{\mu\nu}(t_1,t_2)
=-i \bra{0} {\rm T}\left[ a^\mu_{jH}(t_1) {\bar a}^\nu_{kH}(t_2)\right] \ket{0}
\, ,
\ee
and can be rewritten in the interaction picture as
\be
G_{jk}^{\mu\nu}(t_1,t_2)
=-i \bra{0} {\rm T}\left[{\hat S}a^\mu_j(t_1) {\bar a}^\nu_k(t_2)\right] \ket{0}\,,
\label{eq:TFDDyson}
\ee
because of $\bra{0} {\hat S}^{-1}= \bra{0}$ due to Eq.~(\ref{eq:bra1hatHI}), where ${\hat S}$
was defined in Eq.~(\ref{eq:Smatrix}) with Eq.~(\ref{eq:defV}). Because of the Dyson's
 expansion formula in 
Eq.~(\ref{eq:TFDDyson}) and the existence of the $\xi$-operators, the Wick theorem holds.
Therefore the Feynman diagram method is available in calculating the causal Green's functions
in nonequilibrium TFD.

\section{Self-consistent renormalization condition}

According to the Feynman diagram method, we have the Dyson-Schwinger equation
for the two-point Green's function,
\be
G_{jk}^{\mu\nu}(t_1,t_2)= \Delta_{jk}^{\mu\nu}(t_1,t_2)+
\sum_{j',k'}\int^{\infty}_{-\infty}\int^{\infty}_{-\infty}
\!  \Delta_{jj'}^{\mu\mu'}(t_1,s_1)
\Sigma_{j'k'}^{\mu'\nu'}(s_1,s_2)G_{k'k}^{\nu'\nu}(s_2,t_2)\, ds_1ds_2
\,,
\label{eq:DysonScwinger}
\ee
where the unperturbed propagator is
\be
\Delta_{jk}^{\mu\nu}(t_1,t_2)
=-i \bra{0} {\rm T}\left[a^\mu_j(t_1) {\bar a}^\nu_k(t_2)\right] \ket{0} \, .
\label{eq:unpertPropagator}
\ee
We also define the unperturbed and full propagators for the $\xi$-operators by
\begin{align}
	d^{\mu\nu}_{jk}(t_1,t_2) &= -i \bra{0} {\rm
 T}[\xi^{\mu}_j(t_1) \, {\bar\xi}^{\nu}_{k}(t_2)] \ket{0} \,,\\  
	g^{\mu\nu}_{jk}(t_1,t_2) &= -i \bra{0} {\rm
 T}[\xi^{\mu}_{jH}(t_1) \, {\bar\xi}^{\nu}_{kH}(t_2)] \ket{0} \,, 
\label{eq:xipropagator}
\end{align}
respectively.  These propagators are related to those for the $a$-operators as
\begin{align} \label{eq:Delta}
	\Delta^{\mu\nu}_{jk} (t_1,t_2) &= B_j^{-1, \mu\mu'}(t_1)  
	d^{\mu'\nu'}_{jk}(t_1,t_2) B_{k}^{\nu'\nu}(t_2)\,,\\
	G^{\mu\nu}_{jk} (t_1,t_2) &=  B_j^{-1, \mu\mu'}(t_1)  
	g^{\mu'\nu'}_{jk}(t_1,t_2) B_{k}^{\nu'\nu}(t_2)\,. \label{eq:Gg}
\end{align}
While the unperturbed propagator $d$ has a diagonal structure with respect 
to the thermal index, 
\be
	d^{\mu\nu}_{jk}(t_1,t_2) = \delta_{jk} \BM
	-i\theta(t_1-t_2) & 0 \\ 0 & i\theta(t_2-t_1) \EM^{\mu\nu}
	e^{-i\int_{t_2}^{t_1}\! ds\; \omega_j(s)}\,, 
\ee
the full propagator $g$ has an upper triangular structure in general,
that is, $g^{11}_{jk}(t_1,t_2) \propto \theta(t_1-t_2)$\,,\,$g^{22}_{jk}(t_1-t_2)
\propto \theta(t_2-t_1)$\,,\,$g^{12}_{jk}(t_1,t_2) \ne 0$ and $g^{21}_{jk}(t_1,t_2)
= 0$ \cite{AIP}. This is because $\xi_{jH}^\dagger$ 
and ${\tilde \xi}_{jH}^\dagger$
identically annihilate the bra-vacuum, see Eqs.~(\ref{eq:bra1hatHI}) and (\ref{eq:LTSIalpha}). 
It follows from the tilde-conjugation that $g^{\mu\nu}_{jk}$ has the following properties:
\begin{equation}
g_{jk}^{11}(t_1,t_2)=g_{kj}^{22,\ast}(t_2,t_1) \, , \qquad
g_{jk}^{12}(t_1,t_2)=- g_{kj}^{12,\ast}(t_2,t_1) 
\, .
\end{equation}

The self-energies, defined through the Dyson--Schwinger equations $\Sigma=\Delta^{-1}-G^{-1}$ 
and $S=d^{-1} -g^{-1}$, have the following properties,
\begin{align}
	\Sigma^{\mu\nu}_{jk} (t_1,t_2) &= B_j^{-1, \mu\mu'}(t) 
	S^{\mu'\nu'}_{jk}(t_1,t_2) B_{k}^{\nu'\nu}(t')\,,\\
	\label{S_property3}
	S^{\mu\nu}_{jk}(t_1,t_2) &= 
	\BM S^{11}_{jk}(t_1,t_2) & S^{12}_{jk}(t_1,t_2) \\ 0 &
	S^{22}_{jk}(t_1,t_2)\EM^{\mu\nu} \,,
\end{align}
\vspace{-0.6cm}
\begin{alignat}{2}
	\label{S_property1}
	S^{11}_{jk}(t_1,t_2) &\propto \theta(t_1-t_2) \,,&\qquad
	S^{22}_{jk}(t_1,t_2) &\propto \theta(t_2-t_1) \,,\\
	\label{S_property2}
	S^{11}_{jk}(t_1,t_2) &= S^{22,\ast}_{kj}(t_2,t_1) \,,&\qquad
	S^{12}_{jk}(t_1,t_2) &= - S^{12,\ast}_{kj}(t_2,t_1) \,.
\end{alignat}

We move to perform the self-consistent renormalization, that is, to fix
the two parameters $\omega_j(t)$ and $n_j(t)$ in the unperturbed Hamiltonian $H_{\mathrm u}(t)$.

First recall the on-shell renormalization condition in equilibrium case which is a standard
prescription in ordinary quantum field theory and fixes 
the renormalized energy $\omega_j$.
Then the self-energy depending only on $t_1-t_2$, $S_{jk}^{\mu\nu}(t_1-t_2)$\,, 
is given in $k_0$-representation by 
\be \label{k0-rep}
	S_{jk}^{\mu\nu}(k_0) = \int_{-\infty}^\infty d\tau\; S_{jk}^{\mu\nu}(\tau) e^{ik_0\tau} \,,
\ee
and we have the spectral representation in equilibrium case \cite{AIP},
which becomes in $t$-representation, 
\be
	\Sigma_{jk}^{\mu\nu}(\tau) = 
	\int_{-\infty}^\infty\!\! d\kappa\; B^{-1}[n(\kappa)]^{\mu\mu'} \BM -i\theta(\tau) & 0 \\ 0 & i\theta(-\tau) \EM^{\mu'\nu'}
	B[n(\kappa)]^{\nu'\nu}\, e^{-i\kappa\tau} \, \sigma_{jk}(\kappa)\,,
\ee
with the equilibrium distribution $n(\kappa) = 1/(e^{\beta\kappa} - \sigma)$ for the inverse temperature $\beta$, and 
the spectral function $\sigma_{jk}(\kappa)$.
The $j$-diagonal elements in $k_0$-representation have the properties of
\begin{align}
	& \label{eq:S11S12_equilibrium} S^{11}_{jj}(k_0) = S^{22,\ast}_{jj}(k_0) \,,\\
	&\label{S12_equilibrium}
	S^{12}_{jj}(k_0) = 2i\sigma\pi \bigl( n(k_0) - n_j \bigr) \sigma_{jj}(k_0)\,,
\end{align}
and identically $S^{21}_{jj}(k_0)=0$  as in Eq.~(\ref{S_property3}). 
Because $n_j$ is the equilibrium distribution, 
$S^{12}_{\ell\ell}(k_0)$ automatically vanishes at on-shell: $k_0 = \omega_j$.
Finally we require the on-shell renormalization condition,
\be \label{eq-on-shell}
	0=\Real\; S_{jj}^{11}(\omega_j) = \Real\int_{-\infty}^\infty d\tau\; S^{11}_{jj}(\tau)
 e^{i\omega_j\tau} \,,
\ee
to determine $\omega_j$. Equation (\ref{eq-on-shell}) implies the condition 
$\Real\; S_{jj}^{22}(\omega_j)=0$ as well 
because of Eq.~(\ref{eq:S11S12_equilibrium}). This way all the elements of 
the self-energy $S_{jj}^{\mu\nu}(\omega_j)$ become zero, 
except for the imaginary parts of 
$S_{jj}^{11}(\omega_j)$ and $S_{jj}^{22}(\omega_j)$ representing thermal instabilities.

The self-consistent renormalization condition in the equilibrium case can not be
 extended straightforwardly to the nonequilibrium case.  
Firstly, since the time-dependence of the self-energy is not simply $t_1-t_2$
 but it depends on $t_1$ and $t_2$ separately, its $k_0$-representation and the definition of
the on-shell are not trivial. Secondly, an additional condition is required 
to determine ${\dot n}_j(t)$. For this Chu and Umezawa has proposed their
 renormalization condition not on the self-energy but on the propagator, 
called diagonalization condition \cite{Chu2, Chu1, Chu3}.
The diagonalization condition is 
\be \label{DiagonalCond}
	g_{jj}^{12}(t,t) = 0\,,
\ee
and determines the temporal evolution of $n_j(t)$ in the leading order.
To show it, we solve formally the Dyson--Schwinger equation for $g^{12}$, and obtain
\be \label{Dyson_g12}
	g_{jk}^{12}(t_1,t_2) = \sum_{j'k'}\int\!\! ds_1 d_2 \; 
	g_{jj'}^{11}(t_1, s_1) S_{j'k'}^{12}(s_1, s_2) g_{k'k}^{22}(s_2, t_2) \,.
\ee
Next, we approximate the full propagators $g^{11\,(22)}$ in Eq.~(\ref{Dyson_g12}) 
by the unperturbed ones $d^{11\,(22)}$ 
and divide the self-energy $S$ into a loop contribution $S_{\mathrm{loop}}$ 
and a contribution of the thermal counter term $S_Q$,
\be
	S_{jk,Q}^{\mu\nu}(t_1,t_2) =-i\sigma\Dot{n}_j(t_1)
	\delta_{jk}\delta(t_1-t_2)\, \BM 0 & 1\\ 0 & 0 \EM^{\mu\nu}
	\,. 
\ee
Then we have
\be
	g^{12}_{jj}(t, t) = -i \int_{-\infty}^t\!\!\!ds\; \Biggl[
	\sigma\Dot{n}_j(s) 
	-2\Imag \int_{-\infty}^{s} \!\!\!ds' \; S_{jj, \mathrm{loop}}^{12}(s, s') \,
	e^{i\int_{s'}^s \!ds''\; \omega_j(s'')}\Biggr] \,, 
\ee
and the diagonalization condition Eq.~(\ref{DiagonalCond}) implies the
following transport equation
\be \label{QTE_Diagonal_Cond}
	\Dot{n}_j (t) = 2\sigma\Imag \int_{-\infty}^{t} \!\!\!ds \;
	S_{jj, \mathrm{loop}}^{12}(t, s)\,
	e^{i \int_{s}^t \!ds'\;\omega_j(s')} \,,
\ee
which determines the temporal evolution of $n_j(t)$.
The loop contribution  $S_{jj, \mathrm{loop}}^{12}$ can be 
calculated diagrammatically.

Although the transport equation, obtained by the diagonalization condition (\ref{DiagonalCond}), 
is reduced in the Markovian limit to the ordinary quantum 
Boltzmann equation derived in the other methods, there is a critical problem, that is, 
the diagonalization condition in higher orders leads to 
be inconsistent with the equilibrium theory.  
To confirm it, we express $g^{12}_{jj}(t,t)$ in terms of the unperturbed number distribution
 $n_j(t)$ and the Heisenberg number distribution $n_{j H}(t)$. The latter is defined by
\be
	n_{jH}(t) = \bra{0} {\check a}_{jH}^\dagger(t) {\check a}_{jH}(t) \ket{0}\,,
\ee
or equivalently 
\be
	G^{11}_{jj}(t,t+0) = -i \sigma n_{jH}(t) \,,\qquad
	G^{11}_{jj}(t,t-0) = -i(1+\sigma n_{jH}(t))\,.
\ee
On the other hand,  we obtain from Eq.~(\ref{eq:Gg})
\be
	G^{11}_{jj}(t,t+0) = -i \sigma n_{j}(t) -g^{12}_{jj}(t,t+0)\,,\qquad
	G^{11}_{jj}(t,t-0) = -i(1+\sigma n_{j}(t)) - g^{12}_{jj}(t,t-0)\,,
\ee
and therefore
\be
	g^{12}_{jj}(t,t) = i\sigma (n_{jH}(t) - n_j(t))\,.
\ee
Thus, if the diagonalization condition is applied, the Heisenberg number distribution 
has to be equal to the unperturbed one. But it is not true in the equilibrium limit, where
the full propagator has the spectral representation,
\be
	G_{jk}^{\mu\nu}(k_0) = 
	\int_{-\infty}^\infty\!\! d\kappa\; \left[B^{-1}[n(\kappa)]
	\frac{\rho_{jk}(\kappa)}{k_0-\kappa+i \varepsilon \tau_3}
	B[n(\kappa)]\right]^{\mu\nu} \,,
\ee
with $\displaystyle \int_{-\infty}^\infty\!\!
	 d\kappa\; \rho_{jk}(\kappa) = \delta_{jk}$ and
\be
 n_{jH}= \int_{-\infty}^\infty\!\!  d\kappa\; n(\kappa)\rho_{jj}(\kappa)\, ,
\ee 
and we have $n_{jH}\neq n(\omega_j) $ in general.

We propose a new self-consistent renormalization condition here. For this, the on-shell
energy renormalization condition of the equilibrium Eq.~(\ref{S12_equilibrium})
 has to be extended to the nonequilibrium case, and the diagonalization
 condition Eq.~(\ref{DiagonalCond}) has to be replaced with a new one which 
yields the correct equilibrium limit. It is desirable from a unified viewpoint 
that the two conditions have similar expressions, in contrast to the situation
that the diagonalization condition is imposed on the propagator while
the energy renormalization is the condition on the self-energy. We attempt conditions
on the self-energy.

A natural extension of the on-shell self-energy as in Eq.~(\ref{k0-rep})
 with $k_0 = \omega_j$ to 
the nonequilibrium case with the time-dependent $\omega_j(t)$ would be
\begin{align}
S_{jj}^{\mu\nu}[\omega_j; t] &= 
\int_{-\infty}^\infty d\tau\; S^{\mu \nu}_{jj}(t_1,t_2)
e^{i\int_{t_2}^{t_1} \!ds \; \omega_j(s)}  \,, \nn \\
&= \int_{-\infty}^\infty d\tau\; \left\{ \theta(\tau) S^{\mu \nu}_{jj,r}(t_1,t_2)
+ \theta(-\tau) S^{\mu \nu}_{jj,a}(t_1,t_2)\right\}\,
e^{i\int_{t_2}^{t_1} \!ds \; \omega_j(s)}
\label{eq:S_onshell_noneq}
\end{align}
with $\tau=t_1-t_2$.  There is an ambiguity in this expression, namely, as to
how $t_1$ and  $t_2$ depend on $t$ and $\tau$. Let us parameterize
\be 
t_1= t+r_\ell \tau \,, \qquad t_2= t -(1-r_\ell) \tau \qquad (\ell= r,a)\,,
\ee
for retarded and advanced parts separately, where $r_\ell$ are undetermined parameters.
We expect that $\omega_j(t)$ and ${\dot n}_j(t)$ are determined from 
$S_{jj}^{\mu\nu}[\omega_j; t]$, and the integrands in (\ref{eq:S_onshell_noneq}) should not
involve the quantities later than $t$ from the viewpoint of the thermal causality 
as was already discussed, which implies that $t \geq t_1\, , t_2$ and therefore
$0 \geq r_\ell \tau$ and $ 0\geq -(1-r_\ell) \tau$. We have the consistent choice,
$r_r =0$ for the retarded case $\tau >0$ and $r_a=1$ for the advanced case $\tau <0$.
Thus we acquire the unique expression
\begin{align}
\lefteqn{S_{jj}^{\mu\nu}[\omega_j; t] =}\nn \\ 
& \int_{-\infty}^\infty d\tau\; \left\{ \theta(\tau) S^{\mu \nu}_{jj,r}(t,t-\tau)
e^{i\int_{t-\tau}^{t} \!ds \; \omega_j(s)}
+ \theta(-\tau) S^{\mu \nu}_{jj,a}(t+\tau,t)
e^{i\int_{t}^{t+\tau} \!ds \; \omega_j(s)}\right\}\,.
\end{align}
Now we propose here a new set of renormalization condition, a successor
 to the diagonalization condition (\ref{DiagonalCond}), as
\be
	\label{eq:NewRenoCond}
	S^{12}_{jj}[\omega_j; t] = 0 \,,\quad\mbox{and} \quad
	\Real\,S^{11}_{jj}[\omega_j; t] = 0 \,.
\ee
With the aid of Eq.~(\ref{S_property2}), we obtain
\be
	S_{jj}^{12}[\omega_j; t] = 2i\Imag \int_{-\infty}^t ds \; S^{12}_{jj}(t,s) e^{i\int_{s}^{t} \!ds' \; \omega_j(s')}\,,
\ee
which with the on-shell renormalization condition (\ref{eq:NewRenoCond}) implies the same transport equation as Eq.~(\ref{QTE_Diagonal_Cond}).
Note however that it never means the two conditions (\ref{DiagonalCond}) and (\ref{eq:NewRenoCond}) are equivalent. 
The leading order replacements $g^{11(22)} \to d^{11(22)}$ in the left-hand side of Eq.~(\ref{Dyson_g12}) 
were necessary to derive the transport equation under the diagonalization condition.
While the diagonalization condition gives a correct transport equation only in the leading order, 
our new on-shell renormalization condition is valid in any order.

\section{Summary and discussions}

In this paper it was shown in the superoperator formalism 
that the three basic requirements on thermal field theory
restrict the structure of unperturbed representation in the interaction picture uniquely,
that is, the corresponding unperturbed density matrix has the geometrical number
distribution Eq.~(\ref{eq:GeoDis}). From this representation follows 
 nonequilibrium TFD using the thermal vacuum
and time-dependent Bogoliubov matrix, though it has been an assumption in the nonequilibrium
TFD.

The three basic requirements are 
(a) the existence of quasiparticle picture at each instant of time, 
(b) the thermal causality that the macroscopic quantities should affect the microscopic motions
only in the future but not in the past, 
(c) the relaxation to the equilibrium after a long time.
While the requirement (c) is the thermodynamical law, 
(b) is necessary to provide a consistent causal description in 
the coexistence of microscopic and macroscopic quantities.
The representation space or the particle picture for the nonequilibrium system 
of quantum field with change in the background number distribution is made clear
due to the requirement (a). From the viewpoint of the loop calculation, 
the use of the time-dependent unperturbed superstate (or the unperturbed density matrix)
provides the better approximation than that of time-independent one, usually fixed
at the initial time, because the former takes account of the temporal change in the
number distribution. In connection with (a), we point out that the unperturbed
Liouville--von Neumann equation (\ref{eq:unLvN}) with Eq.~(\ref{eq:Hu3}) is a special
form of the Lindblad equation \cite{Lindblad}.  The Lindblad equation is derived for a
system interacting with an environment system after some coarse-graining in the time 
direction is performed. We consider that following our time-dependent unperturbed
density matrix involves coarse-graining process.

Since the unperturbed representation is written in terms of the two unspecified parameters,
the unperturbed number distribution $n_j(t)$
 and the excitation energy $\omega_j(t)$,  the expectations
of the Heisenberg operators can not be obtained at this stage. The way to fix the parameters 
is known and simple in the equilibrium case, namely,  $n_j$ 
is nothing but the equilibrium distribution $1/(e^{\beta\omega_j}-\sigma)$, 
and $\omega_j$ is interpreted as the renormalized excitation energy and
is determined through the on-shell renormalization condition (\ref{eq-on-shell}). 
It is non-trivial to extend the definition of the on-shell in the nonequilibrium case 
where the time translational invariance is broken. 
Chu and Umezawa proposed the diagonalization condition 
which was not on the self-energy but to on the propagator, 
and derived the quantum transport equation in the leading order.
They have not given an explicit method to determine $\omega_j(t)$, 
and it was approximated by the bare energy.
We have pointed out in this paper that the diagonalization condition becomes inconsistent in the equilibrium limit in higher orders
and have proposed a new renormalization condition which is a natural extension of the on-shell 
renormalization in the vacuum and equilibrium theory
and determine $n_j(t)$  as well as $\omega_j(t)$ in any higher order.  
The concept of the thermal causality removes the ambiguity in defining the on-shell in nonequilibrium situation.

Our discussions throughout this paper are restricted to the cases where there is no spontaneous
breakdown of symmetry. When a symmetry is spontaneously broken and a Bose--Einstein condensate
is formed, we introduce
the Bogoliubov transformation to diagonalize the unperturbed Hamiltonian having terms 
such $\check{a}_j \check{a}_j$. There are various intriguing phenomena 
in the cold atomic gas systems associated with the broken symmetry typified by the unstable condensate
 decay and the quantum phase transition.  We know the two methods to deal with the spontaneous
breakdown of symmetry in nonequilibrium TFD. The first method is to introduce
a $4\times 4$-matrix transformation to combine the thermal and 
usual $2\times 2$-transformations \cite{Elmfors}. But then there are so many parameters
in the  $4\times 4$-matrix that we have not succeeded in extracting physical parameters
yet, as we have extracted the two physical parameters $\omega_j(t)$ and $n_j(t)$ in the
$2\times 2$-matrix. The second method is to expand the field operator in the complete set
of time-dependent wave functions affected by the time-dependent order parameter
\cite{Matsumoto1,NakamuraYamanaka1}. In this method, the $4\times 4$-matrix is given
as a direct product of the two $2 \times 2$-matrices. Both of the methods are not yet
derived from the superoperator formalism, and the extension of the discussions in this paper
to the symmetry broken cases is a future task.

\section*{Acknowledgements}
The authors thank the Yukawa Institute for Theoretical Physics at Kyoto University 
for offering us the opportunity to discuss this work during the YITP workshop 
YITP-W-12-12 on ``Thermal Quantum Field Theories and Their Applications''.
One of the authors (Y.Y.) would like to thank Dr.~K.~Yuasa for discussions on the Lindblad
equation.



\begin{thebibliography}{99}
\bibitem{NRT}
H.~Narnhofer, M.~Requardt, and W.~Thirring, Commun.~Math.~Phys. 92 (1983) 247.

\bibitem{Landsman}
N.P.~Landsman, Ann.~Phys.~(N.Y.) 186 (1988) 141.

\bibitem{CTP}
J.~Schwinger, J.~Math.~Phys. 2 (1961) 407;
L.~V.~Keldysh, Sov.~Phys. JETP 20 (1965) 1018;
L.~P.~Kadanoff and G.~Baym, Quantum Statistical Mechanics, Benjamin, New York, 1962;
P.~Danielewiecz, Ann.~ Phys. (N.Y.) 152 (1984) 239;
K.~Chou, Z.~Su, B.~Hao and L.~Yu, Phys.~Reports 118 (1985) 1.

\bibitem{AIP}
H.~Umezawa, 
Advanced Field Theory --- Micro, Macro, and Thermal Physics, AIP, New York, 1993. 

\bibitem{Chu2}
H.~Chu and H.~Umezawa, 
Int.~J.~Mod.~Phys. A10 (1994) 1693.

\bibitem{Cornell}
M.~H.~Anderson, J.~R.~Ensher, M.~R.~Matthews, C.~E.~Wieman, and E.~A.~Cornell, 
Science 269 (1995) 198.

\bibitem{Ketterle}
K.~B.~Davis, M.~-O.~Mewes, M.~R.~Andrews, N.~J.~van Druten, D.~S.~Durfee, D.~M.~Kurn, and W.~Ketterle, 
Phys.~Rev.~Lett. 75 (1995) 3969.

\bibitem{Bradley}
C.~C.~Bradley, C.~A.~Sackett, J.~J.~Tollett, and R.~G.~Hulet, 
Phys.~Rev.~Lett. 75 (1995) 1687.

\bibitem{NSMOY}
Y.~Nakamura, T.~Sunaga, M.~Mine, M.~Okumura, and Y.~Yamanaka,
Ann.~Phys.~(N.Y.) 325 (2010) 426.

\bibitem{NakamuraYamanaka1}
Y.~Nakamura and Y.~Yamanaka, 
Ann.~Phys.~(N.Y.) 326 (2011) 1070.


\bibitem{Stoof}
H.~T.~C.~Stoof, 
Phys.~Rev.~A 45 (1992) 8398.

\bibitem{Zaremba}
E.~Zaremba, T.~Nikuni, and A.~Griffin,
J.~Low.~Temp.~Phys. 116 (1999) 277.

\bibitem{Gardiner2}
C.~W.~Gardiner, P.~Zoller, R.~J.~Ballagh, and M.~J.~Davis, 
Phys.~Rev.~Lett. 79 (1997) 1793.


\bibitem{Raman}
C.~Raman, M.~Ko\"hl, R.~Onofrio, D.~S.~Durfee, C.~E.~Kuklewicz, Z.~Hadzibabic, W.~Ketterle, 
Phys.~Rev.~Lett. 83 (1999) 2502. 

\bibitem{Schmutz}
Z.~Schmutz, Z.~Phys. B30 (1978) 97.

\bibitem{Henning}
P.~Henning and H.~Umezawa,
Nucl.~Phys. B417 (1994) 463.

\bibitem{Dzhioev}
A.A.~Dzhioev and D.S.~Kosov, 
J.~Phys.: Condens.~Matter 24 (2012) 225304.

\bibitem{Chu3}
H.~Chu and H.~Umezawa, 
Int.~J.~Mod.~Phys. A9 (1994) 2363.

\bibitem{Chu1}
H.~Chu and H.~Umezawa, 
Int.~J.~Mod.~Phys. A9 (1994) 1703.


\bibitem{Lindblad}
G.~Lindblad, Commun.~Math.~Phys. 48 (1976) 119; V.~Gorini, A.~Kossakowski and E.C.G.~Sudarshan,
J.~Math.~Phys. 17 (1976) 821.

\bibitem{Elmfors}
P.~Elmfors and H.~Umezawa, 
Physica A202 (1994) 577.

\bibitem{Matsumoto1}
H.~Matsumoto and S.~Sakamoto, 
Prog.~Theor.~Phys. 105 (2001) 573.




\end{thebibliography}
\end{document}